\documentclass{article}[12pt]
\usepackage{amssymb,amsmath,amsthm}

\def\P{{\bf P}}
\def\ff{\mathcal{F}}
\def\ofp{(\Omega,\ff,\P)}
\def\R{{\mathbb R}}
\def\Z{{\mathbb Z}}
\def\N{{\mathbb N}}

\def\1{{\mathbb 1}}

\def\one{{\bf 1}}

\newtheorem{theorem}{Theorem}[section]
\newtheorem{lemma}[theorem]{Lemma}

\renewenvironment{proof}[1]
{\noindent{\bf Proof.}\hspace{0.1cm} #1} {$\;\blacksquare$\newline}

\makeatletter

\@addtoreset{equation}{section}

\title{\vskip 8.0truecm 
Fluctuations of the Empirical Entropies of a Chain of Infinite
Order \thanks{This work was partially supported by FAPESP (Projeto
Tem\'atico {\it Rhythmic patterns, parameter setting and language
change}, grant 98/3382-0) and Projeto Tem\'atico {\it Critical
phenomena in evolutive processes and equilibrium systems}, grant
(99/11962-9) and Pronex Project {\it Critical phenomena in probability
and stochastic processes} (grant 66.2177/1996-6)}}
\author
{Davide Gabrielli 
\thanks{Work partially supported by FAPESP
(grant 98/11899-2) and Cofin MIUR 2002 (prot. 2002027798$\_$005)} \\
Universit\'a Dell'Aquila
\and 
Antonio Galves
\thanks{Work partially supported by CNPq (grant
301301/79)} \\ 
Universidade de S\~ao Paulo 
\and 
Daniela Guiol 
\thanks{Work supported by a CAPES PhD grant} \\
Universidade de S\~ao Paulo} 
\date{October 28, 2003}

\begin{document}

\maketitle

\begin{abstract}
  This paper addresses the question of the fluctuations of the
  empirical entropy of a chain of infinite order. We assume that the
  chain takes values on a finite alphabet and loses memory
  exponentially fast.  We consider two possible definitions for the
  empirical entropy, both based on the empirical distribution of
  cylinders with length $c\log{n}$, where $n$ is the size of the
  sample and $c$ is a suitable constant. The first one is the
  conditional entropy of the empirical distribution, given a past with
  length growing logarithmically with the size of the sample. The
  second one is the rescaled entropy of the empirical distribution of
  the cylinders of size growing logarithmically with the size of the
  sample.  We prove a central limit theorem for the first one. We also
  prove that the second one does not have Gaussian fluctuations. This
  solves a problem formulated in Iosifescu (1965).
\end{abstract}

\vfill\eject

\section{Introduction}

This paper considers the following question. Suppose we have a
long string of symbols produced by a chain of infinite order. How
does the empirical entropy estimated from the sample fluctuate
around the true entropy of the chain?

By a chain of infinite order, we mean a stationary stochastic process
in which at each step the probability governing the choice of a new
symbol depends on the entire past. We will assume that this dependency
on the past decreases exponentially fast. We will also assume that the
symbols belong to a finite alphabet.

We consider two definitions for the empirical entropy. The first one
is the conditional entropy of the empirical distribution, given a past
with length growing logarithmically with the size of the sample. The
second one is the renormalized entropy of the empirical distribution
of cylinders.  In this case also the size of the cylinders entering in
the definition of the empirical entropy grows logarithmically with the
size of the sample.

In his 1965 classical article, Iosifescu considered the second
definition, in the simpler case in which the length of the cylinders
remains constant and does not grow with the size of the sample. In
this case, he proved that a central limit theorem holds. He observed
at the end of the paper that the problem remained open when the size
of the cylinders is an increasing function of the size of the
sample. That is the situation we consider here. We prove that in this
case, the empirical entropy defined in the second way does not have
Gaussian fluctuations around the theoretical entropy of chain.

In opposition to this negative result, we prove that the central limit
theorem holds for the empirical conditional entropy. These two
theorems are the main results of this paper.

Chains of infinite order seem to have been first studied by Onicescu
and Mihoc (1935) who called them \emph{chains with complete
connections} (\emph{cha\^{\i}nes \`a liaisons compl\`etes}).  Their
study was soon taken up by Doeblin and Fortet (1937) who proved the
first results on speed of convergence toward the invariant
measure. The name chains of infinite order was coined by Harris
(1955).  We refer the reader to Iosifescu and Grigorescu (1990) for a
complete survey, and to Fern\'andez, Ferrari and Galves (2001) for an
elementary presentation of the subject from a constructive point of
view.

Fluctuations of empirical entropies were already studied in Basarin
(1959) for sequences of independent random variables and Markov
Chains. For chains of infinite order, Iosifescu (1965) proved a
central limit theorem for the density of entropy of the empirical
k-marginals when $k$ is fixed and does not change with $n$, the size
of the sample.  The convergence of the empirical entropy when $k$
increases with $n$ was proved in Ornstein and Weiss (1990) (see also
Shields 1996). For a presentation of the problem of the estimation of
the entropy from a physical point of view we refer the reader to
Schurmann and Grassberger (1996).

To prove our theorems we consider the following strategy. We
decompose the difference between the empirical and the theoretical
entropies as the sum of the relative entropy between the empirical
and theoretical marginals plus a Birkhoff sum of some cylindric
functions and a remainder.  It turns out that the rate of
convergence is different for each one of the empirical entropies
we consider.  The first one converges as the conditional entropy
of the marginals of the source, while the second one converges as
the density of entropy of the marginals which is much slower. The
basis of both proofs is a Central Limit Theorem for cylinder
functions with supports of increasing lengths. This theorem is
interesting by itself. Its proof uses a regenerative construction
of the chain of infinite order.

The paper is organized as follows.  Notation, definitions and the
statement of the two main theorems are given in section
\ref{defi}. The proof of the Central Limit Theorem is given in
section 3. In section 4 we prove the asymptotic normality of the
fluctuations of the empirical conditional entropy. Finally, in
section 5 we prove that the density of entropy of the empirical
marginals cannot have Gaussian fluctuations except in the case of
a sequence of independent and identically distributed random
variables.

\section{Definitions and main results}
\label{defi}

Let $A$ be a finite alphabet. Given $x:=\left(x_i\right)_{i\in \Z}\in
A^{\Z}$ and two integers $m\leq n$, the finite sequence
$\left(x_m,...,x_n\right)$ of elements of $A$ will be denoted $x_m^n$.
We will use also the symbol $x_m^n$ with $m$ or $n$ or both
infinite. A sequence $x_{-\infty}^0:=\left(x_i\right)_{i\leq 0}$ will
be called a {\sl history}. Given two histories $x_{-\infty}^{0}$ and
$y_{-\infty}^{0}$, we say that
$x_{-\infty}^0\stackrel{m}{=}y_{-\infty}^0$, if $x_i=y_i$ for all
$i=0,-1, \dots,-m$.

Let $X:=\left(X_i\right)_{i\in\Z}$ be a stationary process taking values
on the finite alphabet $A$ and defined on a probability space
$\ofp$. We denote $\nu$ the law of the chain. This is the unique
measure defined on $A^{\Z}$ such that,
\[
\nu ([a_1^k])= {\bf P}(X_1^k=a_1^k)\, ,
\]
for any $k$ and any sequence $a_1^k$, where $[a_m^n]$ denotes the
cylinder set
\[
[a_m^n]:=\{x\in A^{\Z}: x_m^n=a_m^n\}. \nonumber
\]

Let $\nu_k$ be the marginal of size k of $\nu$. It is a
probability measure on $A^k$ defined by
\[
\nu_k (a_1^k)=\nu ([a_1^k])\, .
\]
We will also use the shorthand
\[
\nu_{k+1}(a_0|a_{-k}^{-1})={\bf P}(X_0=a_0 |
X^{-1}_{-k}=a^{-1}_{-k})\, .
\]
The extensive entropy of order $k$ is defined as
\begin{equation}
H({\nu}_k):= -\sum_{a_1^k\in A^k} \nu_k(a_1^k) \log \nu_k(a_1^k)\
. \nonumber
\end{equation}
The conditional entropy of order $k$ is defined as
\begin{equation}
h(\nu_{k+1}) :=  -\sum_{a_{-k}^0\in
A^{k+1}}\nu_{k+1}(a_{-k}^0)\log\nu_{k+1}(a_0|a_{-k}^{-1})\, ,
\nonumber
\end{equation}
for $k\geq 1$ and
as
\begin{equation}
h(\nu_{1}) :=  -\sum_{a_{0}\in A}\nu_{1}(a_{0})\log\nu_{1}(a_0)\,
, \nonumber
\end{equation}
 for $k=0$.
The entropy of the chain is defined as
\[
 h(\nu) := \lim_{k\to\infty}{\frac{1}{k}H({\nu}_k)}=
 \lim_{k\to\infty}h(\nu_{k})\, .
\]
The fact that the two limits in the definition coincide is a well
known result (see for instance Shields 1996).
Whenever the reference measure $\nu$ is clearly indicated by the context
we will use the short notation
\begin{equation}
H_k=H(\nu_k)\, ,\, h_k= h(\nu_{k+1})\, ,\, h= h(\nu)\, .\label{short}
\end{equation}

In this paper we are concerned with the estimation of $h$, given a
sample of the chain. Let $x_1^n\in A^{n}$ be a finite sample of
the chain $X=\left(X_i\right)_{i\in\Z}$ and take $ k\leq n$. The
empirical k-distribution, given the sample, is the measure on $A^k$
defined as
\[
\hat{\nu}_k(a_1^k;x_1^n):= \frac{1}{n-k+1}\sum_{i=1}^{n-k+1}
{\bf 1}({x_i^{i+k-1}}=a_1^k)\, , \label{empio}
\]
for all $a_1^k\in A^k$, where ${\bf 1}({x_i^{i+k-1}}=a_1^k)$ denotes
the indicator function.
We will also use the short notation
\[
\hat{\nu}_{k,n}(\cdot )=\hat{\nu}_k(\cdot ;x_1^n).
\]
The extensive empirical entropy of order $k$ is defined as
\[
\hat{H}_{k,n} := H(\hat{\nu}_{k,n})\, ,
\]
and the conditional empirical entropy of order $k$ is defined as
\[
\hat{h}_{k,n} := h(\hat{\nu}_{k+1,n}).
\]

We will assume that the process $\left(X_n\right)_{n \in
\Z}$ satisfies the following conditions
\begin{itemize}
\item[(A)] For all $a_1^n\in A^n$
\[
\nu_n (a_1^n) >0.
\]
\item[(B)] The limit
\[
\lim_{m\to\infty}\nu_{m+1} (a_0|a_{-m}^{-1}) := \nu(
a_0|a_{-\infty}^{-1})
\]
exists for all $a_{-\infty}^0\in A^{-\N}$.
\item[(C)] There is a sequence $(\gamma_m)_{m\in\N}$ with $\gamma_m \searrow 0$
when $m \to + \infty$ such that
\[
\sup_{\{a,b\ :\ a_{-\infty}^0\stackrel{m}{=}b_{-\infty}^0\}}
\left|\frac{\nu (a_0|a_{-\infty}^{-1})} {\nu
(b_0|b_{-\infty}^{-1})}-1 \right|=\gamma_{m}.
\]
\end{itemize}

We remark that for i.i.d. random variables $\gamma_0=0$ and for Markov
chains of order $k\ge 1$ condition C is satisfied with
$\gamma_{m}=0$, for all $m\ge k$. We will call $\left(X_n\right)_{n
\in \Z}$ a {\sl chain of infinite order} if $\gamma_{m}>0$ for any
$m$. Whenever $\gamma_{m}\leq Me^{-cm}$ where $M$ and $c$ are suitable
positive constants we will say that the chain loses memory
exponentially fast with rate (at least) $c$. This will be always the
case in this article.

For any measurable function $f:A^{\Z}\to \R$ we will use the
shorthand $\textbf{E}\left[f \right]$ to denote
$\textbf{E}\left[f\left(X_{-\infty}^{+\infty}\right) \right]$. We
will also use $\textbf{E}\left[T_d f\right]$ to denote $\textbf{E}
\left[f\left(T_d X_{-\infty}^{+\infty}\right) \right]$ where $T_d$
denotes the d-step shift $(T_d x)_i=x_{i+d}$, for $d\in\Z$. We
define
\begin{equation}\label{sigma2f}
\sigma^2_f(m):= {\bf E}\left[\left(\frac{1}{\sqrt{m}}
\sum_{d=1}^{m}(T_d f- {\bf E}[f])\right)^2\right].
\nonumber
\end{equation}
A direct computation shows that
\begin{equation}\label{vari}
\sigma^2_f(m)=
C_{f}(0)+2\sum_{d=1}^{m-1}C_{f}(d)-\frac{2}{m} \sum_{d=1}^{m-1}
dC_{f}(d),
\end{equation}
where
\[
C_{f}(d)=\textbf{E}\left[ f T_d f\right] -\textbf{E}\left[ f
\right]\textbf{E}\left[ f \right].
\]

If
\begin{equation}\label{sum}
\sum_{d=1}^{+\infty} dC_{f}(d)<+\infty\, ,
\end{equation}
then
\begin{equation}\label{varilim}
\sigma^2_{f}:=\lim_{m\to \infty}\sigma^2_f(m)=
C_{f}(0)+2\sum_{d=1}^{+\infty}C_{f}(d).
\end{equation}
We recall that a chain of infinite order losing memory exponentially
fast satisfies condition \ref{sum} (cf. \cite{BFG}).

The main results of this paper are the following theorems.

\begin{theorem}\label{ilteorema}
Let $X=\left(X_i\right)_{i\in \Z}$ be a stationary process satisfying
conditions A, B and C and losing memory exponentially fast with rate
$c>\log|A|$. Let us introduce the function $\phi(x):=-\log \nu (x_0
|x_{-\infty}^{-1})$. For any sequence $\left(k(n)\right)_{n\in \N}$ of
positive integers such that $\liminf \frac{k(n)}{\log n} >
\frac{1}{2c}$ and $\limsup \frac{k(n)}{\log n} < \frac{1}{2\log |A|}$
the following statements hold.

$\bullet$ If $\sigma^2_{\phi}>0$, then
\[
\sqrt{\frac{n}{\sigma^2_{\phi}}}
\left(\hat{h}_{k(n),n}- h \right)
\stackrel{{\cal D}}{\to}{\cal N}(0,1)\, ,
\]
where $\stackrel{{\cal D}}{\to}$ means convergence in distribution.

$\bullet$ If $\sigma^2_{\phi}=0$, then
\[
\sqrt{n}\left(\hat{h}_{k(n),n}-h \right)\stackrel{ \bf P}{\to}0\, ,
\]
where $\stackrel{ \bf P}{\to}$ means convergence in
probability.
\end{theorem}

{\noindent{\bf Remark.}
{\it
In Theorem \ref{ilteorema}, if $X$ is a Markov chain of order $m$,
then the lower bound for $k(n)$ becomes simply $\liminf k(n)\geq m$.}

\begin{theorem}\label{negativo}
Let $X=\left(X_i\right)_{i\in\Z}$ be a stationary process satisfying
conditions A, B and C and losing memory exponentially fast with
$\gamma_0>0$ and let $\left(q(n)\right)_{n\in\N}$ be a sequence of
positive real numbers. For any sequence $\left(k(n)\right)_{n\in \N}$
of positive integers such that $k(n)\to +\infty$ when $n\to +\infty$
and such that $\limsup \frac{k(n)}{\log n}< \frac{1}{2\log |A|}$,
the following statements hold.

$\bullet$ If $\displaystyle{\lim_{n\to\infty}\frac{q(n)}{k(n)}=
0}$, then
\[
q(n)\left(\frac{1}{k(n)}\hat{H}_{k(n),n}-h \right) \stackrel{ \bf
P}{\to}0.
\]

$\bullet$ If $\displaystyle{\lim_{n\to\infty}\frac{q(n)}{k(n)}=
\alpha, \ 0<\alpha <+\infty} $, then
\begin{equation}
 q(n)\left(\frac{1}{k(n)}\hat{H}_{k(n),n}-h
\right)\stackrel{\bf P}{\to}\alpha\sum_{i=0}^{\infty}\left(h_i-h
\right)\, ,
\label{fretta}
\end{equation}
where $h$ and $h_i$ are the entropies defined in \ref{short}.

$\bullet$ If $\displaystyle{\lim_{n\to\infty}\frac{q(n)}{k(n)}=
+\infty}$, then for any $r \in \R$
\[
\lim_{n\to\infty}{\bf
P}\left\{q(n)\left(\frac{1}{k(n)}\hat{H}_{k(n),n}-h \right)
< r\right\}=0.
\]
\end{theorem}

{\noindent{\bf Remarks.}
{\it 
Theorem \ref{negativo}  shows that, except for the i.i.d. case, the 
usual $\sqrt{n}$ scaling does not produce asymptotic normality; in 
fact, no scaling does. For the case of i.i.d. processes ($\gamma_0=0$), 
Basarin (1959) \nocite{Bas} proved that the Central Limit Theorem holds 
for the extensive empirical entropy when $q(n)=(n/\sigma^2_{\phi})^{1/2}$.

The theorem remains true if $k(n)\to K<+ \infty$
when $n\to +\infty$, with the only difference that the right hand side
of (\ref{fretta}) is replaced by $\alpha \sum_{i=0}^{K}(h_i-h)$.}

\bigskip

The proofs of Theorems \ref{ilteorema} and \ref{negativo} are based in
the following Central Limit Theorem which is interesting by itself.
Before stating this theorem we need some extra notation.

Given a function
\[
a_{-\infty}^0 \in A^{-\N}\Rightarrow f(a_{-\infty}^0) \in \R
\]
we define its m-variation
as
\[
v_m(f):=\sup_{\{a,b\ :\
  a_{-\infty}^0\stackrel{m}{=}b_{-\infty}^0\}}|f(a)-f(b)|\, .
\]
Given a function $f:A^{\Z}\to \R$ we define its uniform norm as
\begin{equation}
||f ||_{\infty}=sup_{x \in A^{\Z}}|f(x)|\, .
\nonumber
\end{equation}
\begin{theorem}\label{clt}
Let $X=\left(X_i\right)_{i\in\Z}$ be a stationary process satisfying
conditions A, B and C and losing memory exponentially fast, and such
that $\sigma^2_{\phi}>0$, where $\phi$ is defined as in Theorem
\ref{ilteorema}. Let also $\left(f_k\right)_{k\in \N}$ be a family of
uniformly bounded cylindric functions
$f_k(x)=f_k(x_{-k}^0):A^{k+1}\to\R$, satisfying the conditions
\begin{equation}\label{convsup}
\lim_{k \to +\infty}||\phi - f_k ||_{\infty}=0\,
\end{equation}
 and
\begin{equation}\label{varo}
\sup_{m \in \N} v_k\left(f_m\right)\leq M\gamma_k,\ ,
\end{equation}
for all $k \in \N$, where $M$ is a positive constant.  \\
Then for any sequence $\left(k(n)\right)_{n\in \N}$ of positive
integers such that $k(n)\to +\infty$ when $n\to +\infty$ and such
that $\lim_{n\to +\infty} \frac{k(n)^{8}}{ n}=0$, then we have
\begin{equation}
\frac{1}{\sqrt{\sigma^2_{\phi}n}}\sum_{i=1}^n\left\{T_i f_{k(n)}- {\bf
E}\left[f_{k(n)} \right]\right\}
\stackrel{{\cal D}}{\to}{\cal N}(0,1)\, .
\nonumber
\end{equation}
\end{theorem}

We will use the letter $M$ to denote all the constants appearing in
this paper, not only the constant appearing in condition \ref{varo}.

%%%%%%%%%%%%%%%%%%%%%%%%%%%%%%%%%%%%%%%%%%%%%%%%%%%%%%%%%%%

\section{Proof of Theorem \ref{clt}}
\label{cltr}

We start the proof with a lemma on the variances
associated with the functions $f_k$.

%%%%%%%%%%%%%%%%%
\begin{lemma}\label{lemme1}
Let $X=\left(X_i\right)_{i\in\Z}$ be a stationary process and
$\left(f_k\right)_{k\in \N}$ be a sequence of cylindric functions and
let us assume that both satisfy the hypotheses of Theorem
\ref{clt}. Then for any sequence $\left(k(n)\right)_{n\in\mathbb N}$
diverging to $+\infty$ when $n\to\infty$, we have
\[
\lim_{n\to +\infty}\sigma^2_{f_{k(n)}}(n)=\sigma^2_{\phi}.
\]
\end{lemma}

\begin{proof}
Using expressions \ref{vari} and \ref{varilim} we observe that
\begin{eqnarray}\label{ameli}
|\sigma^2_{f_{k(n)}}(n)-\sigma^2_{\phi}| &\leq &
|C_{f_{k(n)}}(0)-C_{\phi}(0)|+
2\sum_{d=1}^{n-1}|C_{f_{k(n)}}(d) -C_{\phi}(d)| \nonumber \\
& & +2\sum_{d=n}^{+\infty}|C_{\phi}(d)|+\frac{2}{n}
\sum_{d=1}^{n-1}d|C_{f_{k(n)}}(d)|\, .
\end{eqnarray}
By direct computation, we have for any $d \ge 0$
\begin{equation}\label{vkn}
|C_{f_{k(n)}}(d) -C_{\phi}(d)|\leq M ||f_{k(n)}-\phi||_{\infty}.
\end{equation}
>From this and the condition \ref{convsup} it follows that
\[
\lim_{n\to \infty }|C_{f_{k(n)}}(0)-C_{\phi}(0)|=0\, .
\]

Let now $\left(g(n)\right)_{n\in \N}$ be a diverging sequence of
positive integers such that
\begin{equation}\label{gn}
\lim_{n\to \infty }g(n)||f_{k(n)}-\phi||_{\infty}= 0\, .
\end{equation}
The second term in the right hand side of \ref{ameli} can be
decomposed as
\begin{equation}\label{dec}
2\sum_{d=1}^{n-1}|C_{f_{k(n)}}(d)-C_{\phi}(d)|
=2\sum_{d=1}^{g(n)}|C_{f_{k(n)}}(d)-
C_{\phi}(d)| +
2\sum_{d=g(n)+1}^{n-1}|C_{f_{k(n)}}(d)-C_{\phi}(d)|
\end{equation}
It follows directly from \ref{vkn} and \ref{gn} that the first term in
the right hand side of \ref{dec} converges to $0$ as $n$ diverges.

To obtain an upper bound for the remaining terms in \ref{ameli} and in
\ref{dec} we use the well known fact that under the conditions of Theorem \ref{clt}
there exist two positive constants $M$ and $\eta$ such that
\[
\left| C_{f_k}(d)\right| \leq M e^{-\eta d} \ \ \mbox{and} \ \ \ \left|
C_{\phi}(d)\right| \leq Me^{-\eta d}\, .
\]
This follows for instance from the main Theorem of \cite{BFG}. This
concludes the proof of the lemma.
\end{proof}

The proof of the Central Limit Theorem uses the fact that under the
hypotheses of Theorem \ref{clt} the chain can be constructed with
a regenerative scheme.  A regenerative scheme is a coupled
construction of the chain $X=\left(X_i\right)_{i\in \Z}$ together with
a stationary renewal process $\left(\tau_i\right)_{i\in \Z}$ in such a
way that the random strings
\[
\left\{\left(X_j\, ;\tau_i \le j <\tau_{i+1}\right)\right\}_{i \in \Z}
\]
are independent and, excepted for $i=-1$, identically
distributed. Here we adopt the convention that
\[
\tau_{-1} \le 0 < \tau_0\, .
\]
The variables $\tau_i$ are called regeneration times of the chain.
The renewal process can be constructed in such a way that the
distance between any two successive regeneration times has all its
moments finite. For a detailed and self-contained presentation of
this type of construction the reader is referred to \cite{FFG}, or
to the original papers \cite{Ber}, \cite{CFF} and \cite{L} .

Given a positive integer $s$, the decimated renewal process
$(\tau^s_i)_{i \in\Z}$ is defined as
\[
{\tau}_{i}^{s}= \tau_{si}\, ,
\] for
every $i \in \Z$. In what follows we will take $s=s(n)$ such that
\begin{equation}
\lim_{n\to \infty}\frac{k(n)^2}{s(n)}=0 \ \hbox{and} \ \lim_{n\to
\infty}\frac{s(n)^4}{n}=0\, , \label{chiocciole}
\end{equation}
this is possible for the hypothesis on the sequence $(k(n))_{n\in
\N}$. From now on, whenever there is no danger of confusion, we
will write $s$ instead of $s(n)$ and $k$ instead of $k(n)$.  We
define
\[
S_{i}(n) := \sum_{j=\tau_i^s+k}
^{\tau_{i+1}^{s}-1} T_{j}f_k-{\bf
E}\left[\sum_{j=\tau_i^s+k}
^{\tau_{i+1}^{s}-1} T_{j}f_k \right]\ .
\]
This definition makes sense when $s(n) \geq k(n)+1$ which is always
the case for $n$ big enough. We recall that with the regenerative
scheme the random strings
\[
\left\{\left(X_j\, ;\tau_i^s \le j <\tau^s_{i+1}\right)\right\}_{i \in \N}
\]
are independent and identically distributed. As a consequence the mean
zero random variables $\left(S_i(n)\right)_{i \in \N}$ are independent
and identically distributed. Using the Berry-Esseen's inequality we
will show that they satisfy the Central Limit Theorem.

We rewrite
\begin{equation}
\sum_{i=1}^n\left(T_if_{k} -{\bf
E}\left[f_{k}\right]\right)
=\sum_{i=0}^{N(n)}S_{i}(n)+R(n)\, ,
\label{rew}
\end{equation}
where
\[
N(n):= \sum_{j=1}^{+\infty}\one(\tau_j^s\leq n)\, ,
\]
and $R(n)$ is the remainder.

The remainder can be written as
\[
R(n)=\sum_{i=1}^{N(n)}R_i(n)+Q(n)\, ,
\]
where
\[
R_{i}(n)= \sum_{j=\tau_{i}^{s}} ^{\tau_{i}^{s}+k-1}
\left(T_{j}f_k\right)-{\bf E}\left[\sum_{j=\tau_{i}^{s}}
^{\tau_{i}^{s}+k-1} \left(T_{j}f_k\right)\right]
\]
and $Q(n)$ is the remainder of the remainder.

\begin{lemma}\label{reminder}
Under the hypotheses of Theorem \ref{clt} we have
\[
\frac{R(n)}{\sqrt{ n}}\stackrel{\bf P}\to 0\, .
\]
\end{lemma}
\begin{proof}
When $n$ is big enough also the random variables
$\left(R_i(n)\right)_{i\in \N}$ are i.i.d.. Using Wald's identity
and Chebychev's inequality we have
\begin{equation}\label{wald}
{\bf P}\left(\left| \frac{\sum_{i=1}^{N(n)} R
_{i}(n) }{\sqrt{n}}\right| >\delta \right)  \leq
\frac{{\bf E}\left[N(n)\right] {\bf E}\left[\left(R
_{i}(n)\right)^2\right]} {\delta^2 n}\, .
\end{equation}
To conclude that the right hand side of inequality \ref{wald}
converges to $0$ we first observe that by the renewal property
\begin{equation}\label{rene}
\lim_{n\to\infty}{\bf E}\left[\frac{N(n)s(n)}{n}\right]=
\frac{1}{{\bf E}\left[\tau_{1}-\tau_{0}\right]}>0.
\end{equation}
Therefore there exists a positive constant $M$ such that
\begin{equation}\label{ENn}
{\bf E}\left[N(n)\right]\le M \frac{n}{s(n)}\,
\end{equation}
for all $n$ big enough.
Since the functions $f_k$ are uniformly bounded above there exist a
positive constant $M$ such that
\begin{equation}\label{Rn}
 {\bf E}\left[\left(R
_{i}(n)\right)^2\right]\le M k(n)^2\, .
\end{equation}
Putting together inequalities \ref{ENn} and \ref{Rn} we conclude that
the right hand side of inequality \ref{wald} is bounded above by
$M\frac{k(n)^2}{\delta^2s(n)}$ where $M$ is a suitable positive
constant.  One of the conditions satisfied by the sequences $k(n)$ and
$s(n)$ implies that this upper bound converges to $0$ as $n$
diverges.

The proof that
\[
\frac{Q(n)}{\sqrt{ n}}\stackrel{\bf P}\to 0\, ,
\]
is analogous. This concludes the proof of the lemma.
\end{proof}

To conclude the proof of the Theorem we must show that
\begin{equation}\label{last}
\frac{\sum_{i=0}^{N(n)}S
_{i}(n) }{\sqrt{\sigma^2_{\phi} n}}\stackrel{\cal D}\to
{\cal N}(0,1).
\end{equation}
The law of large numbers implies that
\[
\frac{N(n)}{\left[{\bf E}\left[N(n)\right]\right]}\stackrel{\bf P}\to 1\, ,
\]
where $\left[{\bf E}\left[N(n)\right]\right]$ denotes the integer part
of the expectation ${\bf E}\left[N(n)\right]$. Therefore, by a
standard argument (see, for instance Theorem 17.1 from Billingsley's
classical treatise \cite{Bil}), \ref{last} will follow from
\begin{equation}\label{lastb}
\frac{\sum_{i=0}^{\left[{\bf E}\left[N(n)\right]\right]}S
_{i}(n) }{\sqrt{\sigma^2_{\phi} n}}\stackrel{\cal D}\to
{\cal N}(0,1).
\end{equation}
To prove \ref{lastb}, before using the Berry-Esseen's inequality we
need to show that the scaling factor appearing in the statement of the
Theorem has the right asymptotic behavior. This is the content of the
next lemma.

%%%%%%%%%%%%%%%%%%%%%%%%%%%%%%%%%%%%%%%%%%%%%%%%%%%%%%%%%%%%%%%%
\begin{lemma}\label{scale}
With the conditions of Theorem \ref{clt} the following
limit holds
\begin{equation}\label{scalefor}
lim_{n \to +\infty}
\frac{{\bf Var}\left[\sum_{i=0}^{\left[{\bf
E}\left[N(n)\right]\right]}
S_{i}(n) \right]}{\sigma^2_{\phi} n}=1\, .
\end{equation}
where ${\bf Var}\left[\cdot \right]$ denotes the variance.
\end{lemma}
\begin{proof}
Using the fact that $S_i(n)$ are mean zero independent random
variables together with \ref{rene}, it is easy to see that
expression \ref{scalefor} is equivalent to
\begin{equation}\label{varanc}
lim_{n \to +\infty} \frac{{\bf
E}\left[\left(S_1(n)\right)^2\right]} {s(n){\bf
E}\left[\tau_{1}-\tau_{0}\right]}=\sigma^2_{\phi}\, .
\end{equation}
To prove \ref{varanc}, we first recall that Lemma \ref{lemme1} assures that
\begin{equation}\label{recallemma1}
\sigma^2_{\phi}=\lim_{n\to +\infty}\sigma^2_{f_{k(n)}}(n)\, .
\end{equation}
Using \ref{rew}, we rewrite the expression inside the limit in the
right-hand side of \ref{recallemma1} as
\begin{equation}\label{lunga2}
l_n\frac{{\bf E}\left[\left(S_1(n)\right)^2\right]}{
s(n){\bf E}\left[\tau_{1}-\tau_{0}\right]}\Big(1+\mathcal{R}(n)\Big)
+\frac{{\bf E}\left[R(n)^2\right]}{n}\, ,
\end{equation}
where
\begin{equation}
\mathcal{R}(n)=
\frac{ 2{\bf E}\left[\left(\sum_{i=0}^{N(n)}
S_{i}(n)\right)R(n)\right]} {{\bf E}\left[N(n)+1 \right] {\bf
E}\left[\left(S_1(n)\right)^2\right]}\, ,
\end{equation}
and $l_n$ is a sequence converging to 1 when $n$ diverges.
We observe that
\[
\lim_{n \to +\infty}\frac{{\bf E}\left[R(n)^2\right]}{n}=0\,
\]
A proof of this fact, ignoring the remainder $Q(n)$, is given in Lemma
\ref{reminder}. The remainder can be treated in a similar way.

Using the Cauchy-Schwartz inequality and
Wald's identity we obtain the upperbound
\begin{equation}\label{boundfin}
\left|\mathcal{R}(n)\right| \leq M\sqrt{\frac{{\bf
E}\left[\left(R_1(n)\right)^2\right]} {{\bf
E}\left[\left(S_1(n)\right)^2\right]}}\leq \frac{Mk(n)}{
\sqrt{{\bf E}\left[\left(S_1(n)\right)^2\right]}}\, ,
\end{equation}
where $M$ is a suitable positive constant.

If
\[
\lim_{n \to +\infty} \frac{k(n)}{
\sqrt{{\bf E}\left[\left(S_1(n)\right)^2\right]}}=0\, ,
\]
then the proof would be concluded. Let us assume that this is not the
case. Then for an $\epsilon > 0$ there exist infinitely many
values of $n$ such that
\[
  k(n)> \epsilon\sqrt{{\bf E}\left[\left(S_1(n)\right)^2\right]}
\, .
\]
However using this inequality inside \ref{lunga2} would led to a
result which is in contradiction with Lemma \ref{lemme1}. This
concludes the proof of the lemma.
\end{proof}

%%%%%%%%%%%%%%%%%%%%%%%%%%%%%%%%%%%%%%%%%%%%%%%%%%%%%%%%%%%%
To conclude the proof of Theorem \ref{clt} we will show that
\begin{equation}\label{goal}
\lim_{n \to +\infty} \sup_{r\in \R}\left|{\bf P} \left\{
\frac{\sum_{i=0}^{\left[{\bf E}\left[N(n)\right]\right]}S _{i}(n)
}{\sqrt{\sigma^2_{\phi} n}}\leq r\right\}-\Psi (r)\right|=0 \, ,
\end{equation}
where
\[\Psi(r)=\frac{1}{\sqrt{2\pi}}\int_{-\infty}^re^{-u^2/2}du\, .\]
By Lemma \ref{scale}
\begin{equation}
\frac{\sum_{i=0}^{\left[{\bf E}\left[N(n)\right]\right]}S _{i}(n)
}{\sqrt{\sigma^2_{\phi} n}}= l_n\frac{\sum_{i=0}^{\left[{\bf
E}\left[N(n)\right]\right]}S _{i}(n) }{\sqrt{{\bf E}\left[\left(S
_{1}(n)\right)^2\right]\left[{\bf E}\left[N(n)+1\right]\right]}},
\nonumber
\end{equation}
where $l_n \to 1$, when $n\to + \infty$. Therefore the left hand side
of expression \ref{goal} can be rewritten as
\begin{equation}\label{goal2}
 \sup_{r\in \R}\left|{\bf P} \left\{\frac{\sum_{i=0}^
 {\left[{\bf E}\left[N(n)\right]\right]}S
_{i}(n) }{\sqrt{{\bf E}\left[\left(S
_{1}(n)\right)^2\right]\left[{\bf E}\left[N(n)+1\right]\right]}}
\leq r/l_n \right\}-\Psi (r)\right|
\end{equation}
We modify the expression inside the absolute value of \ref{goal2}, by
 adding and subtracting $\psi(r/l_n)$ and then using the triangle's
 inequality. The second term of the upper bound obtained this way is
 just
\[
\sup_{r \in \R}|\Psi(r)-\Psi(r/l_n)|
\]
which goes to $0$ as $n$ diverges.  Now we finally use the
Berry-Esseen Theorem in the remaining term to obtain the upper bound
\begin{equation}\label{ub}
 \sup_{r\in \R}\left|{\bf P} \left\{\frac{\sum_{i=0}^ {\left[{\bf
 E}\left[N(n)\right]\right]}S _{i}(n) }{\sqrt{{\bf E}\left[\left(S
 _{1}(n)\right)^2\right]\left[{\bf E}\left[N(n)+1\right]\right]}} \leq
 r/l_n \right\}-\Psi (r/l_n)\right| \le \frac{M{\bf E}\left[
 \left|S_1(n)\right|^3\right]} {\left({\bf E}\left[\left(S
 _{1}(n)\right)^2\right] \right)^{\frac{3}{2}}\sqrt{\left[{\bf E}
 \left[N(n)+1\right]\right]}}
\end{equation}
We refer the reader to the classical treatise of Shiryaev
\cite{Sh} for a presentation of the Berry-Esseen Theorem.

The fact that the functions $f_{k(n)}$ are uniformly bounded provides
an upper bound for the numerator of \ref{ub}
\begin{equation}\label{ub1}
{\bf E}\left[
\left|S_1(n)\right|^3\right] \le Ms(n)^3 \, ,
\end{equation}
where $M$ is a positive constant. From \ref{rene} we obtain the lower
bound
\begin{equation}\label{lb1}
\left[{\bf E}\left[N(n)+1\right]\right] \ge M \frac{n}{s(n)}\, ,
\end{equation}
where $M$ is another positive constant. Finally as a byproduct of Lemma
\ref{scale} we obtain the lower bound
\begin{equation}\label{lb2}
{\bf E}\left[\left(S _{1}(n)\right)^2\right] \ge M s(n)\, ,
\end{equation}
where $M$ is a third positive constant. Putting together the bounds \ref{ub}
\ref{ub1}, \ref{lb1} and \ref{lb2} we conclude that
\begin{equation}
 \sup_{r\in \R}\left|{\bf P} \left\{\frac{\sum_{i=0}^
 {\left[{\bf E}\left[N(n)\right]\right]}S
_{i}(n) }{\sqrt{{\bf E}\left[\left(S
_{i}(n)\right)^2\right]\left[{\bf E}\left[N(n)+1 \right]\right]}}
\leq r/l_n \right\}-\Psi (r/l_n)\right| \le
M\frac{s(n)^2}{\sqrt{n}}\, . \nonumber
\end{equation}
By hypothesis this last upper bound goes to $0$ as $n$ diverges, and
this concludes the proof of the Theorem.

\section{Proof of theorem \ref{ilteorema}}

We first consider the case $\sigma^2_{\phi}>0$.  Let us introduce the
cylinder functions $\phi_k(x)=\phi_k(x_{-k}^0)$ defined by
$\phi_k(x):=-\log \nu_{k+1}(x_0|x_{-k}^{-1})$ and define

\[
D(n)=\sqrt{n}\left(\hat{ h}_{k(n),n}-h\right)-
\frac{1}{\sqrt{n}}
\sum_{i=1}^{n}\left(T_{i}\phi_{k(n)}(x)-\textbf{E}\left[\phi_{k(n)}\right]\right)
\, .
\]
It is easy to see that the sequence of functions $(\phi_k)_{k \ge
1}$ satisfies the conditions of Theorem \ref{clt}. Therefore if
$\sigma^2_{\phi}>0$, it follows that
\[
\frac{1}{\sqrt{\sigma^2_{\phi}n}}\sum_{i=1}^n\left\{T_i \phi_{k(n)}- {\bf
E}\left[\phi_{k(n)} \right]\right\}
\stackrel{{\cal D}}{\to}{\cal N}(0,1)\, .
\]
Therefore to prove Theorem \ref{ilteorema} it is enough to show that
\begin{equation}\label{zero}
D(n)\stackrel{P}{\to} 0\, ,
\end{equation}
as $n$ diverges. To prove \ref{zero} we will decompose $D(n)$ in three
parts and show that each one of them converges to $0$ in probability. The
parts are defined as follows
\[
D_1(n)=\sqrt{n}\left(h_{k(n)}-h \right)\, ,
\]
\[
D_2(n)=\sqrt{n}\left[
H(\hat{\nu}_{k(n),n}|\nu_{k(n)})-
H(\hat{\nu}_{k(n)+1,n}|\nu_{k(n)+1})\right]\, ,
\]
where
\[
H(\hat{\nu}_{k,n}|\nu_{k})=\sum_{a_1^k}\hat{\nu}_{k,n}(a_1^k)\log
\left[\frac{\hat{\nu}_{k,n}(a_1^k)} {\nu_k(a_1^k)}\right]\, ,
\]
and
\[
D_3(n)=\frac{\sqrt{n}}{n-k(n)}
\sum_{i=k(n)+1}^{n}T_{i}\phi_{k(n)}(x)-\frac{1}{\sqrt{n}}
\sum_{i=1}^{n}T_{i}\phi_{k(n)}(x)\, .
\]
The expert reader has already identified $H(\hat{\nu}_{k,n}|\nu_{k})$
as the relative entropy between the probability measures
$\hat{\nu}_{k,n}$ and $\nu_k$ on $A^k$.

\begin{lemma}\label{lemmad1}
Let $X$ be a stationary chain satisfying conditions A, B and C and
losing memory exponentially fast with rate at least $c$, where $c$ is
any fixed positive real number. For any sequence $(k(n))_{n\in \N}$ of
positive integers such that
\begin{equation}\label{conditiononk}
\liminf_{n \to +\infty} \frac{k(n)}{\log n}>\frac{1}{2c}
\end{equation} we have
\begin{equation}
\lim_{n \to +\infty}D_1(n)=0\, .
\nonumber
\end{equation}
\end{lemma}

\begin{proof}
By Jensen's inequality
\begin{equation}\label{modulo}
h_k-h \geq 0\, .
\end{equation}
By definition
\begin{equation}\label{mod1}
h_k-h= \int d\nu(a_{-\infty}^{0})\left(
\log\frac{\nu(a_{0}|a_{-\infty}^{-1})}
{\nu_{k+1}(a_{0}|a_{-k}^{-1})}\right).
\end{equation}
By standard calculus
\begin{equation}\label{dk}
\int d\nu(a_{-\infty}^{0})\left(
\log\frac{\nu(a_{0}|a_{-\infty}^{-1})}
{\nu_{k+1}(a_{0}|a_{-k}^{-1})}\right)\le \int d\nu
(a_{-\infty}^{0})\left(
\frac{\nu(a_{0}|a_{-\infty}^{-1})}
{\nu_{k+1}(a_{0}|a_{-k}^{-1})}-1\right)\, .
\end{equation}
It follows from hypothesis C that
\begin{equation}\label{lega}
\int d\nu (a_{-\infty}^{0})\left(\frac{\nu(a_{0}|a_{-\infty}^{-1})}
{\nu_{k+1}(a_{0}|a_{-k}^{-1})}-1\right) \le \gamma_k\, .
\end{equation}
By hypothesis
\begin{equation}\label{gale}
\gamma_k \le M\exp\{-ck\}\, ,
\end{equation}
where $M$ is a strictly positive constant.  Using together expressions
\ref{modulo}, \ref{mod1}, \ref{dk}, \ref{lega} and \ref{gale} we
obtain
\begin{equation}
0 \le D_1(n) \le M\sqrt{n}\exp\{-ck(n)\}\, .
\nonumber
\end{equation}
The conclusion of the lemma now follows directly from \ref{conditiononk}.
\end{proof}

Before proving that $D_2(n)\stackrel{{\bf P}}{\to}0$ we need the
following lemma.

\begin{lemma}\label{carato}
If $X$ is a stationary chain satisfying
conditions A, B , C and losing memory exponentially fast then
\[
\sup_{k,n}\sup_{a\in A^{\Z}}\left(\frac{\sigma^2_{\one
[a_1^{k}]}(n)} {\nu([a_1^{k}])}\right)< +\infty\, ,
\]
where $\one[a_1^{k}]=\one (x_1^k=a_1^k )$ denotes the indicator
function of the cylinder set $[a_1^k]$.
\end{lemma}

\begin{proof}
From \ref{vari} we obtain:
\begin{equation}\label{vari2}
\sigma^2_{\one [a_1^k]}(n) \leq
\left|C_{\one [a_1^{k}]}(0)\right| +2\sum_{d=1}^{n-1}\left(1+\frac{d}{n}\right)
\left|C_{\one [a_1^{k}]}(d)\right|.
\end{equation}
We need an upper bound for the  right hand side of \ref{vari2}.
The $\phi$-mixing property of the chain implies that
\begin{equation}\label{expo}
|C_{\one [a_1^k]}(d)|\leq M\nu ([a_1^k])e^{- \eta d}.
\end{equation}
where $\eta$ and $M$ are suitable positive constants. A
constructive proof of this can be performed using the regenerative
construction which was introduced in the proof of Theorem
\ref{clt} (cf. \cite{FFG}).  To conclude the proof it is enough to
use \ref{expo} in the right hand side of \ref{vari2}.
\end{proof}

\begin{lemma}\label{d2}
Let $X$ be a stationary chain satisfying
conditions A, B and C and losing memory exponentially fast. For
any sequence $(k(n))_{n\in \N}$ of positive integers such that
\begin{equation}\label{conditiononk2}
\limsup_{n \to +\infty} \frac{k(n)}{\log n} < \frac{1}{2\log|A|}
\end{equation} we have
\begin{equation}
D_2(n)\stackrel{{\bf P}}{\to} 0 \, .
\nonumber
\end{equation}
\end{lemma}
\begin{proof}
By definition
\[
H(\hat{\nu}_{k,n}|\nu_{k})=\sum_{a_1^k}\hat{\nu}_{k,n}(a_1^k)\log
\left[\frac{\hat{\nu}_{k,n}(a_1^k)} {\nu_k(a_1^k)}\right]\, .
\]
Therefore by Jensen's inequality
$
0 \leq H(\hat{\nu}_{k,n}|\nu_{k})$.
This allows us to use Markov's inequality to get
\begin{equation}\label{markov}
{\bf P}\left\{ H(\hat{\nu}_{k(n),n}|\nu_{k(n)}) \ge \frac{\delta}{\sqrt{n}} \right\}
\le \frac{\sqrt{n}}{\delta}{\bf E}\left[H(\hat{\nu}_{k(n),n}|\nu_{k(n)})
\right]\, ,
\end{equation}
for any $\delta>0$.

We need an upper bound for the expectation on the right hand side of
inequality \ref{markov}. We use again Jensen's inequality to obtain

\[
H(\hat{\nu}_{k,n}|\nu_{k})
\le
\log{\left\{\sum_{a_1^k}\left[\frac{\left(\hat{\nu}_{k,n}(a_1^k)\right)^2}
{\nu_k(a_1^k)}\right]\right\}}
\]

We observe that
\begin{equation}\label{observe}
\sum_{a_1^k}\left[\frac{\left(\hat{\nu}_{k,n}(a_1^k)\right)^2}
{\nu_k(a_1^k)}\right]=1+ \sum_{a^{k}_{1}}
\frac{(\hat{\nu}_{k,n}(a^{k}_{1}) -\nu_{k}(a^{k}_{1}))^2}
{\nu_{k}(a^{k}_{1})}\, .
\end{equation}
Therefore from \ref{observe} and standard calculus it follows that
\begin{equation}\label{standard}
H(\hat{\nu}_{k,n}|\nu_{k})\, \le\,  \sum_{a^{k}_{1}}
\frac{(\hat{\nu}_{k,n}(a^{k}_{1}) -\nu_{k}(a^{k}_{1}))^2}
{\nu_{k}(a^{k}_{1})}\, .
\end{equation}
Using the definition of $\hat{\nu}_{k,n}$, we can rewrite the right
hand side of \ref{standard} as
\begin{equation}\label{rewrite}
\frac{1}{n-k+1} \sum_{a^{k}_{1}}\frac{1}{\nu_{k}(a^{k}_{1})}
\left\{\frac{1}{\sqrt{n-k+1}}\sum_{i=1}^{n-k+1}\Big({\bf}{\one} (
x_i^{i+k-1}=a^{k}_{1})- \nu_{k}(a^{k}_{1})\Big)
\right\}^2\, .
\end{equation}
Therefore from  \ref{standard} and \ref{rewrite} it follows that
\begin{equation}\label{end}
{\bf E}\left[H(\hat{\nu}_{k(n),n}|\nu_{k(n)}) \right]
\leq  \frac{1}{n-k(n)+1}\
\displaystyle{\sum_{a_1^{k(n)}}}
\frac{\displaystyle{\sigma^2_{\one[a_1^{k(n)}]}(n-k(n)+1)}}
{\displaystyle{\nu_{k(n)} (a_1^{k(n)})}}\, .
\end{equation}
By Lemma \ref{carato} the right hand side of \ref{end} is bounded
 above by
\begin{equation}\label{end2}
\frac{M|A|^{k(n)}}{n-k(n)+1}\, ,
\end{equation}
where $M$ is a positive constant.
Putting together inequalities \ref{markov}, \ref{end}, \ref{end2}
we conclude that
\[
{\bf P}\left\{H(\hat{\nu}_{k(n),n}|\nu_{k(n)})\ge
\frac{\delta}{\sqrt{n}}\right\} \le
\frac{M|A|^{k(n)}\sqrt{n}}{\delta\left(n-k(n)+1\right)}\, .
\]
By hypothesis \ref{conditiononk2} this last upper bound converges
to $0$ as $n$ diverges. This concludes the proof of the lemma.
\end{proof}

\begin{lemma} From the hypotheses of Lemma \ref{d2} it follows that
\[
D_3(n)\stackrel{{\bf P}}{\to} 0 \, .
\]
\end{lemma}
\begin{proof}
The result follows immediately from the fact that functions
$\phi_{k}$, $k \in \N$ are uniformly bounded.
\end{proof}

This concludes the proof of Theorem \ref{ilteorema} in the case
$\sigma^2_{\phi}>0$.

Let us finally consider the case $\sigma^2_{\phi}=0$. With our
assumptions, the variance vanishes if and only if the chain is an
i.i.d. sequence of random variables with
\[
{\bf P}\left\{X_n=a\right\}=\frac{1}{|A|}\, ,
\]
for any $a \in A$. In this case
\[
\phi_k(x)=\phi(x)=\log|A|\, ,
\]
for any $x$ and all $k$. This implies that
\[
\frac{1}{\sqrt{\sigma^2_{\phi}n}}\sum_{i=1}^n\left\{T_i \phi_{k(n)}- {\bf
E}\left[\phi_{k(n)} \right]\right\}=0
\]
and the conclusion of the second part of Theorem \ref{ilteorema}
follows trivially.

\section{Proof of theorem \ref{negativo}}

We follow the same approach as for the proof of Theorem
\ref{ilteorema}. We define
\begin{equation}\label{decompo}
W(n)=q(n)\left[\frac{1}{k(n)}\hat{H}_{k(n),n}-h\right]-
\frac{q(n)}{n}\sum_{i=1}^{n}\left\{T_i\bar{\phi}_{k(n)}-
{\bf E}[\bar{\phi}_{k(n)}]\right\}\, ,
\end{equation}
where the functions $\bar{\phi}_{k}$ are cylinder functions defined by
\[
\bar{\phi}_{k}:=\frac{1}{k+1}\sum_{i=0}^{k}\phi_k\,
\]
and $\phi_k(x)=\phi_k(x_{-k}^0)$ are the cylinder functions introduced
in the proof of Theorem \ref{ilteorema} and defined by
$\phi_k(x):=-\log \nu_{k+1}(x_0|x_{-k}^{-1})$, when $k\ge 1$, and
$\phi_0(x):=-\log \nu_{1}(x_0)$.

We will decompose $W(n)$ in three parts defined as follows
\[
W_1(n)=q(n)\left(\frac{1}{k(n)}H_{k(n)}-h\right)\, ,
\]
\[
W_2(n)=-\frac{q(n)}{k(n)}H(\hat{\nu}_{k(n),n}|\nu_{k(n)})\, ,
\]
and
\[
W_3(n)=W(n)- W_1(n)-W_2(n)\, .
\]

Let us briefly sketch the proof. The functions $\bar{\phi}_k$ satisfy
the conditions of Theorem \ref{clt} and by assumption we have
$\sigma^2_{\phi}>0$. Therefore if we take $q(n)=\sqrt{n}$, then the
second term of right hand side of \ref{decompo} converges to a normal
distribution with variance $\sigma^2_{\phi}$. At this point we could
try to reproduce the proof of Theorem \ref{ilteorema} and to show that
the remainder $W(n)$ vanishes as $n$ diverges.  This would imply that
the first term of right hand side of \ref{decompo} is asymptotically
normal.

The second and the third terms of the remainder both converge to
$0$ in probability, when $q(n)=\sqrt{n}$. However, with this
choice of $q(n)$, the term $W_1(n)$ diverges to $+\infty$.

That's the end of the history. There is no scaling factor which
simultaneously assures asymptotic normality for the second term in the
right-hand side of \ref{decompo} and convergence to $0$ for the
remainder term.  This impossibility is due to the slow rate of
convergence of the extensive entropy, much slower than the rate for
the conditional entropy.  This is content of the next lemmas. In all
of them we will assume that $X$ is a chain satisfying the hypotheses
of Theorem \ref{negativo}.

\begin{lemma}\label{w1}
For any sequence $\left(k(n)\right)_{n\in \N}$
of positive integers such that $k(n)\to +\infty$ when $n\to +\infty$
the following statements hold.

$\bullet$ If $\displaystyle{\lim_{n\to\infty}\frac{q(n)}{k(n)}=
0}$, then
\[
\lim_{n\to+\infty} W_1(n)=0\, .
\]

$\bullet$ If $\displaystyle{\lim_{n\to\infty}\frac{q(n)}{k(n)}=
\alpha, \ 0<\alpha <+\infty} $, then
\[
\lim_{n\to+\infty} W_1(n)=\alpha\sum_{i=0}^{+\infty}(h_i-h)\, ,
\]
with $ 0<\sum_{i=0} ^{+\infty}(h_i-h)<+\infty$.

$\bullet$ If $\displaystyle{\lim_{n\to\infty}\frac{q(n)}{k(n)}=
+\infty}$, then
\[
\lim_{n\to+\infty} W_1(n)=+\infty\, .
\]
\end{lemma}
\begin{proof}
By definition, we have that
\[
W_1(n)=\frac{q(n)}{k(n)}\sum_{i=0}^{k(n)-1}(h_i-h)\, .
\]
Moreover, Jensens inequality implies that $\sum_{i=0}
^{+\infty}(h_i-h)\ge 0$.  This is actually a strict inequality, since, by assumption, $\gamma_0>0$ and therefore the chain $X$ is not
a sequence of independent random variables.  The fact that $\sum_{i=0}
^{+\infty}(h_i-h)<+\infty$ follows easily from expressions \ref{mod1},
\ref{dk} and \ref{lega}. The lemma follows immediately from these facts.
\end{proof}

\begin{lemma}\label{w2}
For any sequence $k(n)$ satisfying the hypotheses of
Theorem \ref{negativo} and for any sequence $q(n)$
such that $q(n)\le M k(n)\sqrt{n}$,
with $M$ a positive constant we have that
\[
W_2(n)\stackrel{\P}{\to}0\, ,
\]
as $n$ diverges.
\end{lemma}
\begin{proof}
The proof is identical to the proof of Lemma \ref{d2}.
\end{proof}

\begin{lemma}\label{w3}
There exists a positive constant $M$ such that
\[
W_3(n)\le M\frac{q(n)k(n)}{n}\, .
\]
\end{lemma}
\begin{proof}
The result follows immediately from the fact that the functions
$\bar{\phi}_{k}$, $k\in \N$, are uniformly bounded.
\end{proof}
Lemmas \ref{w1}. \ref{w2} and \ref{w3} together with Theorem
\ref{clt} imply that Theorem \ref{negativo} holds for sequences
$q(n)$ bounded above by $ M\sqrt{n}$, where $M$ is a positive
constant.

\begin{lemma}\label{qgrande}
If $\limsup_{n\to+\infty}q(n)/\sqrt{n}=+\infty$ and $\lim_{n \to
+\infty}q(n)/k(n)=+\infty$, then for any $r>0$ we have
\[
\lim_{n \to +\infty}\P\left\{
q(n)\left(\frac{\hat{H}_{k(n),n}}{k(n)}-h\right) >r\right\}=1\, .
\]
\end{lemma}
\begin{proof} By hypothesis for any positive $M$, we have
$q(n)>M\sqrt{n}$ for infinitely many values of $n$. Let us define
\[
\tilde{q}(n)= \min\{ q(n), M \sqrt{n}\}\, .
\]
Obviously we have
\begin{equation}\label{ovvio}
\P\left\{
q(n)\left(\frac{\hat{H}_{k(n),n}}{k(n)}-h\right) >r\right\}
\ge
\P\left\{
\tilde{q}(n)\left(\frac{\hat{H}_{k(n),n}}{k(n)}-h\right) >r\right\}\, .
\end{equation}
The part already proved of Theorem
\ref{negativo} implies that the right hand-side of inequality
\ref{ovvio} converges to $1$. This concludes the proof of Lemma
\ref{ovvio} as well as the proof of Theorem \ref{negativo}.
\end{proof}

\section{Acknowledgments}
We thank J.R. Chazottes, P. Collet, D. Duarte, R. Fern\'andez,
N.L. Garcia, S. Popov and P. Shields for many interesting discussions
on the subject. The authors thank the ZiF of the University of
Bielefeld, the Laboratoire de Recherches Informatiques de
l'Universit\'e de Paris-Sud and the Di\-par\-ti\-men\-to di Fi\-si\-ca
dell'U\-ni\-ver\-si\-t\`a di Ro\-ma {\sl La Sa\-pienza} for their
hospitality during the preparation of this paper.

\vskip30pt

\vskip30pt

Davide Gabrielli

Dipartimento di Matematica Pura ed Applicata,

Universit\'a Dell'Aquila,

via Vetoio loc. Coppito

67100 L'Aquila,

Italia

e-mail: {\tt gabriell@univaq.it }

\vskip 1cm

Antonio Galves

Instituto de Matem\'atica e Estat\'{\i}stica,

Universidade de S\~ao Paulo,

Caixa Postal 66281,

05315-970 S\~ao Paulo, SP

Brasil

e-mail: {\tt galves@ime.usp.br, dana.rv@bluewin.ch}

\end{document}